\documentstyle[manuscript,epsfig,array,amsmath,prb,aps]{revtex}

\draft

\title{Quasi-Two-Dimensional Melting in Porous Media: Effect of Multi-layers and Cross-Over in Scaling Behavior }

\author{Malgorzata Sliwinska-Bartkowiak$^1$, Ravi Radhakrishnan$^2$ and Keith E Gubbins$^3$}

\address{{$^1$Instytut Fizyki, Uniwersytet im Adama Mickiewicza, Poznan, Poland \\
$^2$Department of Bioengineering, University of Pennsylvania, Philadelphia, PA, USA\\
$^3$Department of Chemical Engineering, North Carolina State University, Raleigh NC, USA.}}

\date{\today}

\begin{document}
\maketitle

\renewcommand{\theequation}{\arabic{equation}}
\renewcommand{\thefigure}{\arabic{figure}}
\setcounter{equation}{0}  
\setcounter{figure}{0}  

\begin{abstract}
We report molecular simulation and experimental results for simple
fluids adsorbed in activated carbon fibers (ACF), where the adsorbed
phase consists of either one or two molecular layers. Our molecular
simulations involve smooth pore-walls for large system sizes and a
systematic system size-scaling analysis. We provide calculations of
the Ginzburg parameter to monitor the self-consistency of our finite
size simulation results, based on which we establish that for system
sizes smaller than 60 molecular diameters, fluctuations are too large
to uphold the finite size simulation results. We present scaling
analysis and free energy results for a system size of 180 molecular
diameters. Our scaling analysis reveals a two-stage melting in our
bilayer system consistent with KTHNY scaling for the order parameter
correlation functions. Based on the Lee-Kosterlitz scaling of the free
energy surface, we establish that the transitions are first order in
the thermodynamic limit. Similar results for a monolayer established
that the transitions were continuous. We provide scaling arguments to
suggest that the change in the order of the transitions in going from
a monolayer to bilayer adsorbed system results from the interactions
of defects in the multilayers. We also report experimental
measurements of CCl$_4$ and aniline (with two confined molecular
layers) adsorbed in ACF. The differential scanning calorimetry and
dielectric relaxation spectroscopy measurements for the transition
temperatures are in quantitative agreement with each other and with
our simulation results. We also report nonlinear dielectric effect
(NDE) measurements and show that our data is consistent with the NDE
scaling law for the KTHNY scenario, which we derive. Finally, we
discuss the significance of the six-fold corrugated potential and the
effect of the strength of pore-wall potential on the melting
behavior. Both the simulations and experiments show that the hexatic
phase is stabilized by confinement in carbon pores; for carbon
tetrachloride and aniline the hexatic phase is stable over a
temperature range of 55~K and 26~K, respectively. Moreover, this
stability increases as the ratio of fluid-wall to fluid-fluid
attraction increases.
\end{abstract}


\section{Introduction}
\label{sec:intro}

For continuous symmetry breaking transitions (such as melting
transitions) in two-dimensions, the Mermin-Wagner theorem states that
true long range order ceases to exist\cite{mermin1966}. Halperin and
Nelson proposed the ``KTHNY''
(Kosterlitz-Thouless-Halperin-Nelson-Young) mechanism for melting of a
crystal in two dimensions\cite{nelson1979}, which involves two
transitions of the Kosterlitz-Thouless~(KT)
kind\cite{kosterlitz1972}. The crystal to hexatic transition occurs
through the unbinding of dislocation pairs, and the hexatic to liquid
transition involves the unbinding of disclination pairs.  Each KT
transition is accompanied by a non-universal peak in the specific heat
above the transition temperature, associated with the entropy
liberated by the unbinding of the vortex (dislocation or disclination)
pairs, and by the disappearance of the stiffness coefficient
associated with the presence of quasi-long-range order in the
system. The KTHNY theory predicts that the correlation function
associated with the translational order parameter in the crystal
decays algebraically with exponent $\eta <1/3$, while long range
orientational order is maintained, and the correlation function
associated with the orientational order parameter in the hexatic phase
decays algebraically with exponent $0 < \eta_6 <1/4$ while there is no
translational order. The KTHNY theory is an analysis of the limit of
stability of a two-dimensional solid, since it neglects the existence
of the liquid phase and thereby does not impose an equality of the
chemical potentials of the solid and liquid phases as the criterion
for melting. Therefore, other pathways for two-dimensional melting can
not be ruled out. For example, it is possible for the dislocation
unbinding transition to be pre-empted by grain-boundary-induced
melting, as shown by the work of Chui\cite{chui1982}, which predicts
that the critical value of the defect core energy ($E_c$) above which
the melting cross-over from grain boundary induced melting to
two-stage KTHNY melting is ~\mbox{$E_c = 2.8 k_BT$}\cite{strandburg1988}.
Excellent reviews are available on the subject of two-dimensional
melting\cite{strandburg1988} (also see section~\ref{sec:previous}).

\paragraph*{}

In this paper we report molecular simulation and experimental results
for simple fluids adsorbed in activated carbon fibers (ACF), whose pore widths are such that they can accommodate one or two molecular layers. Our computer
simulations involve smooth walls for large system sizes, and we perform a
systematic scaling analysis. We provide calculations of the
Ginzburg parameter to monitor the self-consistency of our finite size
simulation results for system sizes (box lengths) up to 180 molecular diameters. We
also present scaling analysis and free energy results for a system
size of 180 molecular diameters. Based on the Lee-Kosterlitz scaling
of the free energy surface, we establish the nature (first order
vs. continuous) of the transitions. We also report experimental
measurements based on differential scanning calorimetry, dielectric
relaxation spectroscopy, and nonlinear dielectric effect for CCl$_4$
and aniline (with two confined molecular layers) adsorbed in ACF.

\section{Previous Work}
\label{sec:previous}

Experimental and computer simulation studies on the subject of
two-dimensional melting often have the underlying objective of
establishing whether or not the pathway of melting conforms to KTHNY
behavior: (1) Does melting occur in two stages mediated by a hexatic
phase? (2) Do the order parameter correlation functions associated
with the crystal and hexatic phases have the appropriate scaling
behavior? (3) Are the observed phase transitions first order or second
order in the thermodynamic limit? The different experimental systems
explored in this regard include \textit{free-standing
liquid-crystalline (LC) films}\cite{brock1989,lin2000},
\textit{confined colloidal
suspensions}\cite{murray1988,marcus1997,karnchanaphanurach2000,zahn1999},
 and \textit{adsorbed fluid on a planar
substrate}\cite{heiney1983,specht1984,mctague1982,rosenbaum1983,greiser1987,motteler1985,kim1986,zhang1986}.

\vspace{\baselineskip}

\noindent \textit{Computer Simulation Studies of Two-Dimensional
Melting.} Early simulation studies on small (\mbox{$\lesssim 10000$
atoms}), strictly two-dimensional systems failed to provide compelling
evidence to support the KTHNY melting
scenario\cite{strandburg1988,undink1987,stan1989,weber1995}. For
systems with repulsive interactions, it was shown by Bagchi
et. al.\cite{bagchi1996}, using a systematic scaling analysis on large
system sizes ~\mbox{($\simeq 64,000$ atoms)}, and subsequently by
Jaster\cite{jaster1999}, that the equilibrium properties are indeed
consistent with the KTHNY theory of melting (a single stage melting
via first-order transitions with finite correlations was ruled out).
Bagchi et. al.\cite{bagchi1996} have also observed a first order
single stage melting for a two-dimensional system of disks interacting
with a soft repulsive potential for a \textit{small system size},
crossing over to \textit{two-stage continuous} melting consistent with
KTHNY scaling for larger system sizes, suggesting the earlier studies
were plagued by serious system size effects.

\paragraph*{}

Computer simulation studies that directly mimic the liquid-crystalline
thin film experiments have not been attempted owing to the complexity
of the molecular architecture and experimental conditions. Idealized
simulations of spheres confined between hard walls and interacting
with each other via a model potential that mimic the screened
potential in colloidal systems\cite{marcus1997} (but otherwise neglect
the solvent) have been performed by Bladon and
Frenkel\cite{bladon1995}, and Zangi and Rice\cite{zangi1998}. The
former study reported a strong dependence (and qualitatively different
phase diagrams of the melting region) of melting behavior on the
parameters of the inter-molecular interaction of a two-dimensional
square-well fluid, and attributed the difference to the changing
values of the defect core energy\cite{bladon1995}; in particular, the
study puts the two-dimensional hard disks in the single-stage first
order melting regime (rather than the KTHNY regime). In subsequent
experiments and simulations carried out by Rice and co-workers using
two different screened (effective) potentials with which confined
colloidal spheres
interact\cite{marcus1997,karnchanaphanurach2000,zangi1998}, a strong
dependence of the melting scenario on the inter-particle interaction,
consistent with the computer simulation study by Bladon and
Frenkel\cite{bladon1995} was reported. A clear picture regarding the
differing scenarios has not yet emerged due to: (a) the experimental
studies on colloidal systems are faced with the question of attainment
of thermodynamic equilibrium\cite{marcus1997,karnchanaphanurach2000},
while the simulation results are for a small system
size\cite{bladon1995}, and are therefore subject to finite-size effects; (b)
the calculated value for the defect core energy $E_c$ for hard disks
lie in the KTHNY regime\cite{sengupta2000}; (c) the qualitative
behavior of these studies in the limit of a hard sphere potential
differs from that of a recent simulation study by
Jaster\cite{jaster1999} on a large 2-d system of hard disks.

\paragraph*{}

Several computer simulations aimed to mimic an adsorbed fluid on
graphite have been
reported\cite{abraham1980,roth1998,alavi1990}. Abraham\cite{abraham1980}
has reported several studies of xenon on graphite in which the melting
transition temperatures are in quantitative agreement with
experiment. However, the existence of an intrinsic hexatic phase could
not be demonstrated by the simulations, which were performed on fairly
small system sizes. Similar conclusions were drawn by
Roth\cite{roth1998} based on simulations of krypton on graphite, again
for a small system size. Alavi\cite{alavi1990} reported a simulation
study of CD$_4$ on MgO (using a small system size) with melting
behavior consistent with the KTHNY scenario and a defect core energy
of $9 k_BT$ (KTHNY regime). One notable difference between the CD$_4$
on MgO system and the Kr, Xe on graphite systems is that, in the
former case, due to the molecular structure of CD$_4$ adsorbed on MgO,
the relevant degrees of freedom at the temperature of study make the
adsorbate move in a field-free manner. 

\paragraph*{}

In summary, the simulations on
small systems have failed to provide compelling evidence of the
melting picture, and therefore can not be used conclusively to fill in
the gaps or interpret experimental measurements. While the system-size
scaling analysis for large system sizes in the simulations of
Bagchi et.~al.\cite{bagchi1996} (which indicated a KTHNY melting scenario)
clearly underlines the need to use large system sizes in studying the
two-dimensional melting problem, even such methods do not provide a
source of distinction between the first-order and continuous nature of the
transition.

\vspace{\baselineskip}

\noindent \textit{Two-dimensional Melting in Adsorbates Confined in
Porous Media.} Activated carbon fibers~(ACF) possess
micro-crystallites made up of graphene sheets that tend to align in
similar directions, with slit shaped voids between the
microcrystals. The spontaneous ordering of the molecules adsorbed in
these voids into distinct two- dimensional molecular layers (analogous
to the structure of a smectic-A phase in liquid crystals) makes the
adsorbed phase a quasi-two-dimensional system. Microporous ACF can be prepared having a range of mean pore sizes, ranging from those that can accommodate just a
single layer of the adsorbate in the micropores to those that can
accommodate a few layers\cite{gelb1999}. The microcrystals in ACF
are themselves arranged as an amorphous matrix; therefore the
maximum in-plane correlation length of the adsorbed fluid is thought
to be limited by the average size of the microcrystals. Real samples
of ACF have different average sizes for their microcystals (ranging
from 5~nm to 100~nm) depending on the method of activation and
the source of the carbon material used to derive the porous
material\cite{gelb1999}. Electron micrographs are commonly employed to
determine the average size of the microcrystals, while nitrogen
adsorption isotherms are used to determine the average pore size and
pore size distribution. The substrate field induced by the porous
matrix is thought to have a six-fold symmetry like that of
graphite. Additionally, due to the high density of carbon atoms in the graphene microcrystals, the confined
fluid feels a large potential well, and this increases in depth for
decreasing pore sizes; this unique feature induces pronounced layering
effects even in multi-layer adsorbate phases, each layer effectively
being two-dimensional. The quasi-two-dimensional phase of adsorbed
simple fluids (spherical molecules) do not have the
complication of additional coupling due to any herring bone symmetry, in contrast to the LC systems. In addition, owing to the nanoscopic length
scales, these systems are thermally driven and can be studied under
equilibrium (a challenging issue for the colloidal systems). The large
value of the defect core energy\cite{alavi1990} in adsorbed systems
and the role of out-of-plane motions in stabilizing the hexatic phase in these
quasi-two-dimensional systems have
been noted earlier\cite{strandburg1988,zangi1998}.

\section{Molecular Simulation Methods}
\label{sec:simmethods}

We performed Grand Canonical Monte Carlo~(GCMC) simulations of a fluid
adsorbed in slit-shaped pores of width $H$, where $H$ is defined as
the perpendicular distance between the planes passing through the
nuclei of the first layer of molecules that make up the pore walls of
the slit-shaped pore. The interaction between the adsorbed fluid
molecules is modeled using the Lennard-Jones~(12,6) potential with
size and energy parameters, $\sigma_{\rm{ff}}, \, \mbox{and} \,
\epsilon_{\rm{ff}}$. The Lennard -Jones potential was cut-off at a
distance of $5 \sigma_{\rm{ff}}$, beyond which it was assumed to be
zero. The pore walls were modeled as a continuum of LJ molecules using
the``10-4-3'' Steele potential\cite{steele1973}, given by,
\begin{equation}
\phi_{\rm{fw}}(z)=2 \pi \rho_{\rm w} \epsilon_{\rm{fw}} \sigma_{\rm{fw}}^2 \Delta \left [
\frac{2}{5} \left (\frac{\sigma_{\rm{fw}}}{z} \right )^{10} - \left
( \frac{\sigma_{\rm{fw}}}{z} \right )^4 - \left ( \frac{\sigma_{\rm{fw}}^4}
{3 \Delta (z+0.61 \Delta)^3} \right ) \right ]
\label{eq:steele}      
\end{equation}
\normalsize 
Here, the $\sigma$'s and $\epsilon$'s are the size and energy
parameters in the Lennard-Jones~(LJ) potential, the subscripts~f and~w
denote fluid and wall respectively, $\Delta$ is the distance between
two successive lattice planes of pore wall, $z$ is the coordinate
perpendicular to the pore walls and $\rho_{\rm w}$ is the number
density of the wall atoms. For a given pore width, $H$, the total
potential energy from both walls is given by,
\begin{equation}
\phi_{\rm{pore}}(z)= \phi_{\rm{fw}}(z) + \phi_{\rm{fw}}(H-z)
\label{eq:porepotential}
\end{equation}         
The strength of attraction of the pore walls relative to the
fluid-fluid interaction is determined by the coefficient $\alpha$,
\begin{equation}
\alpha = \frac{\rho_{\rm w} \epsilon_{\rm{fw}} \sigma_{\rm{fw}}^2
 \Delta}{\epsilon_{\rm{ff}}}
\label{eq:alpha}
\end{equation}
 in equation~\ref{eq:steele}. 
The fluid-fluid interaction parameters were chosen to be
those for CCl$_4$\cite{radhakrishnan2002}, with
$\sigma_{\rm{ff}}=0.514$~nm, $\epsilon_{\rm{ff}} / k_B=395$~K; with these parameters the freezing temperature of bulk CCl$_4$ is reproduced. The
parameters for the graphite pore wall interaction to mimic the
micropores in ACF were taken from Steele\cite{steele1973},
($\sigma_{\rm{ww}}=0.34$~nm, $\rho_{\rm w}=114$~nm$^{-3}$,
$\epsilon_{\rm{ww}} / k_B = 28$~K, and $\Delta=0.335$~nm). The
Lorentz-Berthlot mixing rules, together with the ff and ww parameters,
were used to determine the values of $\sigma_{\rm{fw}}$ and
$\epsilon_{\rm{fw}}$. The simulation runs were performed by fixing the
chemical potential, $\mu$, the volume ,$V$, of the pore and the
temperature, $T$. Two pore-widths $H = 1 \sigma_{ff} + \sigma_{fw} =
0.91$ nm (monolayer of adsorbate) and $H = 2 \sigma_{ff} +
\sigma_{fw} = 1.41$ nm (bilayer of adsorbate) , were chosen for
study. The rectilinear dimension of the cells were therefore $180
\sigma_{\rm{ff}} \times 180 \sigma_{\rm{ff}} \times H$ (93~nm $\times$
93~nm $\times$ $H$~nm). Typically the system consisted of up to 64,000
adsorbed fluid molecules. Periodic boundary conditions were employed
in the $x$ and $y$ directions (the $xy$-plane being parallel to the
pore walls) and no long range corrections were applied. The adsorbed
molecules formed distinct molecular layers parallel to the plane of
the pore walls. The simulation was set up such that insertion,
deletion and displacement moves were chosen at random with equal
probability. The calculations for the larger of the chosen pore-width
also enable a direct comparison to be made with experimental
measurements for CCl$_4$ confined in porous pitched-based activated carbon fiber
ACF~A-10, of mean pore width $H=1.4$~nm. We expect the approximation
of a structureless graphite wall to be a good one here, since the
diameter of the LJ molecule (0.514~nm) is much larger than the C-C
bond length in graphite (0.14~nm). The state conditions were such that
the confined phase was in equilibrium with bulk LJ~CCl$_4$ at
1~atm. pressure. The simulations were started from a well equilibrated
confined liquid phase at $T=400$~K, and in successive simulation runs
the temperature was reduced. Equilibration was for a minimum of 11
billion steps; the standard deviation of block averages of total
energy of the system was $10^{-4}$ and the rate of insertion was equal
to that of deletion to a factor of $10^{-6}$.

\vspace{\baselineskip}

\noindent \textbf{Free Energy Determination.} Motivated by the work of
Frenkel and co-workers\cite{vanduijneveldt1992,lyndenbell1993,radhakrishnan1999}, we
employ the Landau free energy approach that was successful in our
earlier
studies\cite{radhakrishnan2002,lyndenbell1993,radhakrishnan1999,radhakrishnan2002a}. The
Landau-Ginzburg formalism \cite{landau1980} involves choosing a
spatially varying order parameter $\Phi (\vec{r})$, that is sensitive
to the degree of order in the system. We use a two- dimensional bond
orientational order parameter to characterize the orientational order
in each of the molecular layers that is defined as
follows\cite{mermin1968}:
\begin{equation}
\Psi_{6,j}(\vec{\rho}~) \,=\, \frac{1}{N_b} \, \sum_{k=1}^{N_b} \,
\exp(i6 \theta_k),
\label{eq:hexatic}
\end{equation}
where $N_b$ is the number of nearest-neighbor bonds and $\theta_k$ is the bond angle (see below); the summation is over the (imaginary) bonds joining the central molecule to its nearest neighbors. $\Psi_{6,j}(\vec{\rho}~)$ measures the hexagonal bond order at
position $\vec{\rho}~$ in the $xy$ plane within each layer $j$, and is
calculated as follows. Nearest neighbors were identified as those
particles that were less than a cutoff distance~$r_{\rm{nn}}$ away
from a given particle, and belonged to the same layer. We used a
cutoff distance~$r_{nn}=1.3~\sigma_{\rm{ff}}$, corresponding to the
first minimum of $g(r)$. The orientation of the nearest neighbor bond
is measured by the $\theta$ coordinate, which is the angle that the
projection of the nearest-neighbor vector on to the $xy$-plane makes
with the $x$ axis. $\Psi_{6,j}(\vec{\rho}~)$, is calculated using
equation~\ref{eq:hexatic}, where the index $k$ runs over the total
number of nearest neighbor bonds $N_b$ at position $\vec{\rho}~$, in
layer $j$. The order parameter $\overline{\Psi}_{6,j}$ in layer $j$ is
given by ~\mbox{$\overline{\Psi}_{6,j} \, = \, | \int \, d \vec{\rho} \,
\Psi_{6,j}(\vec{\rho}~) | \, / \, \int \, d \vec{\rho}$}. For the case
of LJ~CCl$_4$ in slit-shaped pores, where there is significant
ordering into distinct molecular layers, the spatially varying order
parameter $\Phi(\vec{r}~)$ can be reduced to $\Phi(z)$, and can be
represented by,
\begin{equation}
\label{eq:average}
\Phi(z) \, =\, \sum_{j=1}^{n} \; \overline{\Psi}_{6,j} \, \delta (z - \hat{z}_j)
\end{equation}
In equation~(\ref{eq:average}), the sum is over the number of adsorbed
molecular layers, and $\hat{z}_j$ is the $z$ coordinate of the plane in
which the coordinates of the center of mass of the adsorbed molecules
in layer $j$ are most likely to lie. It must be recognized that
each of the $\overline{\Psi}_{6,j}$'s are variables that can take
values in the range $[0,1]$; the number of layers is $n=1$ or $n=2$ in our
system. The histograms are collected to evaluate the probability
distribution ~\mbox{$P[\overline{\Psi}_{6,1},\overline{\Psi}_{6,2}]$}.

\paragraph*{}

The Landau free energy\cite{landau1980}
$\Lambda[\overline{\Psi}_{6,1},\overline{\Psi}_{6,2}]$ (for the bilayer system) is given by,
\begin{equation}
\Lambda[\overline{\Psi}_{6,1},\overline{\Psi}_{6,2}] \,=\, -k_BT \, \ln (P[\overline{\Psi}_{6,1},\overline{\Psi}_{6,2}]) \, + \, \mbox{Constant}
\label{eq:landau-prob}
\end{equation}
The Landau free energy is computed by a histogram method combined with
umbrella sampling\cite{torrie1974}, using the probability distribution
~\mbox{$P[\overline{\Psi}_{6,1},\overline{\Psi}_{6,2}]$}. Detailed procedures
to collect statistics, construct the histograms and to choose the the weighting functions for performing the umbrella sampling are described
elsewhere\cite{vanduijneveldt1992,lyndenbell1993,radhakrishnan1999}. The
grand free energy of a particular phase~A, ~\mbox{$\Omega_A=-k_BT \,\ln(\Xi)$}
(where $\Xi$ is the partition function in the grand canonical
ensemble), is related to the Landau free energy by,
\begin{equation}
\exp(- \beta \Omega) \,=\,  \prod_{j=1}^2 \int_j \, d
\overline{\Psi}_{6,j} \,
\exp(- \beta \Lambda[\overline{\Psi}_{6,1},\overline{\Psi}_{6,2}])
\label{eq:grand}
\end{equation}
where the limits of integration in equation~\ref{eq:grand} are from
the minimum value of
~\mbox{$(\overline{\Psi}_{6,1},\overline{\Psi}_{6,2})$} to the maximum
value of ~\mbox{$(\overline{\Psi}_{6,1},\overline{\Psi}_{6,2})$} that
characterize the phase~A. The grand free energy is computed via
numerical integration of equation~\ref{eq:grand}. The two-
dimensional, in-plane positional and orientational correlation
functions ($g_j(r)$ and $G_{6,j}(r)$ of layer $j$), were monitored to
keep track of the nature of the confined phase. The positional pair
correlation function is the familiar radial distribution function. The
orientational pair correlation function is
~\mbox{$G_{6,j}(\rho)=\langle \Psi_{6,j}^*(0)\Psi_{6,j}(\rho)
\rangle$}, where ~\mbox{$\vec{\rho} = x \hat{e}_x +y \hat{e}_y$}. The
above equations correspond to the two-layer case. For the one layer
case, the Landau free energy reduces to
$\Lambda[\overline{\Psi}_{6,1}]$ and the index $j$ in
equation~\ref{eq:grand} assumes a value of 1.

\section{Experimental Methods}
\label{sec:exptmethods}

\noindent \textbf{Sample Preparation.} The liquid samples were reagent
grade chemicals, and were distilled twice prior to use in the
experiment. The conductivities of the purified dipolar fluid samples
were found to be less than $10^{-10}$~ohm$^{-1}$m$^{-1}$. The
microporous activated carbon fiber (ACF) samples used were
commercially available from Osaka gas company, Japan, with a pore size
distribution of about $5\%$ around the mean pore
diameter\cite{kaneko_psd}. ACF samples with an average pore width of
1.41~nm were used. The pore samples were previously characterized by
Oshida and Endo by obtaining electron
micrographs\cite{gelb1999,oshida1998}, and by Kaneko and co-workers
using nitrogen adsorption measurements\cite{kaneko_psd}. The
characterization results for ACF showed that these amorphous materials
consisted of uniform pores formed by graphitic microcrystals, with an
average microcrystal size of 7--10~nm\cite{gelb1999}.

\vspace{\baselineskip}

\noindent \textbf{Differential Scanning Calorimetry~(DSC).} A
Perkin-Elmer DSC7 differential scanning calorimeter was used to
determine the melting temperatures and latent heats of fusion, by
measuring the heat released in the melting of aniline and ${\rm{CCl}}_4$.  The
temperature scale of the DSC machine was calibrated using the melting
temperature of pure fluid from the literature. The temperature
scanning rates used for the melting and freezing runs varied from
$0.1$K$/$min to $0.5$K$/$min. The background of each raw DSC spectrum
was subtracted, based on a second-order polynomial fit to the measured
heat flow away from the signals of interest. The melting temperatures
in the bulk and confined systems were determined from the position of
the peaks of the heat flow signals, and the latent heats were
determined based on the scaled area under these signals. The melting
temperature was reproducible to within $0.5$~$^{\mbox{o}}$C for fluid
adsorbed in pores. The latent heats were reproducible to within $5\%$.

\vspace{\baselineskip}

\noindent \textbf{Dielectric Relaxation Spectroscopy~(DRS).} The
relative permittivity of a medium, ~\mbox{$\kappa^* = \kappa_r- i \kappa_i$},
is in general a complex quantity whose real part $\kappa_r$ (also
known as the dielectric constant) is associated with the increase in
capacitance due to the introduction of the dielectric. The imaginary
component $\kappa_i$ is associated with mechanisms that contribute to
energy dissipation in the system, due to viscous damping of the
rotational motion of the dipolar molecules in alternating fields. The
latter effect is frequency dependent. The experimental setup consisted
of a parallel plate capacitor of empty capacitance
$C_o=4.2~\mbox{pF}$. The capacitance, $C$, and the tangent loss, $\tan
( \delta)$, of the capacitor filled with the fluid between the plates
were measured using a Solartron 1260 gain impedance analyzer, in the
frequency range $1~\mbox{Hz~-}~10~\mbox{MHz}$, for various
temperatures. For the case of the adsorbate in ACF, the sample was
introduced between the capacitor plates as a suspension of ground ACF
particles of $0.1~\mbox{mm}$ mesh ACF particles in pure fluid. Due to the
large conductivity of ACF, the electrodes were blocked by teflon. The
relative permittivity is related to the measured quantities by:
\begin{equation}
\kappa_r \,=\, \frac{C}{C_o}; \, \kappa_i \,=\, \frac{\tan (\delta)}
{\kappa_r}
\label{eq:kappa}
\end{equation}
In equation~(\ref{eq:kappa}), $C$ is the capacitance, $C_o$ is the
capacitance without the dielectric and $\delta$ is the angle by which
current leads the voltage. The complex
dielectric permittivity, ~\mbox{$\kappa^*=\kappa_r-i\kappa_i$}, is measured as a
function of temperature and frequency.

\paragraph*{}

For an isolated dipole rotating under an oscillating electric field in
a viscous medium, the Debye dispersion relation is derived using
classical mechanics\cite{debye1929},
\begin{equation}
\kappa^* \,= \, \kappa_{\infty ,r} \, + \, \frac{\kappa_{s,r} -
    \kappa_{\infty ,r}}{1+ i \omega \tau}
\label{eq:debye}
\end{equation}
Here $\omega$ is the frequency of the applied potential and $\tau$ is
the orientational (rotational) relaxation time of a dipolar
molecule. The subscript $s$ refers to static permittivity (low
frequency limit, when the dipoles have sufficient time to be in phase
with the applied field). The subscript $\infty$ refers to the optical
permittivity (high frequency limit) and is a measure of the induced
component of the permittivity. Further details of the experimental
methods are described elsewhere\cite{sliwinska1999,chelkowski1980}.
The dielectric relaxation time (molecular orientational relaxation
time for dipolar molecules) was calculated by fitting the dispersion
spectrum of the complex permittivity near resonance to the Debye model
of orientational relaxation.

\vspace{\baselineskip}

\noindent \textbf{Nonlinear Dielectric Effect~(NDE).} The nonlinear
dielectric effect (NDE) is defined as the permittivity change, $\Delta
\epsilon$, of the medium in a strong electric field, $E$:
\begin{equation}
\rm{NDE} \,=\, \frac{\Delta \epsilon}{E^2} \,=\, \frac{\epsilon_E - \epsilon_{E=0}}{E^2}
\label{eq:nde1}
\end{equation}
where $\epsilon_E$ is the permittivity of the medium in a field
$E$. The permittivity $\epsilon$ is related to the dielectric constant
$\kappa_r$ by the relationship $\kappa_r= \epsilon / \epsilon_o$,
$\epsilon_o$ being the permittivity in vacuum. Note that within linear
response (for weak fields), the permittivity is independent of the
applied electric field; However, for strong fields, the most general
form for the permittivity (consistent with field reversal invariance)
can be written as ~\mbox{$\epsilon_E = \epsilon_{E=0} + b E^2 + c E^4 +
\cdots$}. By definition, NDE in equation~\ref{eq:nde1} represents the
first order (non-linear) response consistent with the general
equation. The NDE was measured using the pulse method using
rectangular millisecond pulses of the electric field with amplitudes
ranging from $4\times 10^7$ to $9 \times 10^7$V/m. The separation
between the invar electrodes in the measuring condenser was $2 \times
10^{-4}$m, and the changes in the capacitance were measured to an
accuracy of $5 \times 10^{-4}$pF \cite{sliwinska-bartkowiak1983}.

\section{Molecular Simulation Results}
\label{sec:molres}

Our simulations were made to mimic two experimental systems of CCl$_4$
confined in ACF with one as well as two confined molecular layers. The orientational
correlation function $G_{6,j}(r)$ for equilibrated liquid, hexatic and
crystal phases are shown in figure~\ref{fig:corr} \marginpar{\fbox{Fig.~\ref{fig:corr}}}. The long-range
orientational order in the crystal phase, the algebraic decay of
orientational order in the hexatic phase and the exponential decay of
orientational order in the liquid phase are captured in the
plots. Since the phases are characterized by long-ranged correlations, we
check for the attainment of equilibrium and artifacts due to finite
size of the simulations as described below.

\paragraph*{}

Simulation results are always for a finite system size, and
system-size effects can not be avoided. When the state conditions are
such that the relevant order-parameter correlation length $\xi$
approaches or exceeds the spatial extent of the simulation cell $L$,
the system crosses over to a mean-field
behavior\cite{privman1990,mon1992}. For correlations in the
bond-orientational order parameter, this is the case in the hexatic
phase at all temperatures, and in the liquid phase near the
liquid-hexatic transition. Under these circumstances, the self
consistency of the mean field result is checked by calculating the
Ginzburg parameter~$\gamma_{\rm{GL}}$\cite{chaikin1995},
\begin{equation}
\gamma_{\rm{GL}} \,=\, \frac{{ \langle \overline{\Psi}}_{6,j}^2 \rangle}
{\langle{ \overline{\Psi}}_{6,j}\rangle^2} \,=\, \frac{I(L)}{\langle{\overline{\Psi}}_{6,j}\rangle^2 } \, - 1.
\label{eq:ginzparam}
\end{equation}
where $L^2 \times H=V$, the volume of the system, $L$ is the length of the simulation box, and $I(L)$ is given by
\begin{equation}
I(L) = \frac{\int_{V} \, d \vec{\rho} \, G_{6,j} (\rho)}
{\int_{V} \,d \vec{\rho}}
\label{eq:ginzparam2}
\end{equation}
The Ginzburg parameter gives the ratio of the variance in the order
parameter (due to thermal fluctuations) to the square of the average value of the
order parameter. Mean field results are clearly suspect in the regime
~\mbox{$\gamma_{\rm{GL}} \gtrsim 1$}, when long-range fluctuations are likely to destroy
the observed order in the finite size simulations. If ~\mbox{$\gamma_{\rm{GL}} \ll 1$}, the simulations are
self-consistent, and the ordered phase observed in the simulations is
stable against thermal fluctuations. The Ginzburg parameter for
different system sizes is calculated (using equation~\ref{eq:ginzparam}) by numerically integrating
equation~\ref{eq:ginzparam2}, and are reported in Table~I. \marginpar{\fbox{Tab~I}} Our calculations
for the bilayer system ($H=1.41$~nm) indicate that the simulation results for these quasi-two-dimensional
systems are only reliable for systems sizes ~\mbox{$\gtrsim 60
\sigma_{\rm{ff}}$}. For smaller system sizes, the bond-orientational
order parameter fluctuations are too large to validate the finite size
results from computer simulations. Therefore, we report the free
energy calculations and scaling behavior for system sizes of~60 and~$180 \sigma_{\rm{ff}}$.

\paragraph*{}

To quantify the scaling properties of the orientational correlation
function (figure~\ref{fig:corr}) in each phase, we plot ~\mbox{$\log
[I(L_B)/I(L)]$} as a function of $\log[L_B/L]$ in
figure~\ref{fig:corr2}, \marginpar{\fbox{Fig.~\ref{fig:corr2}}} for different values of the block
length. The block length $L_B < L$ defines the subsystem $L_B \times
L_B \times H$ for which the integrals (equation~\ref{eq:ginzparam2})
were evaluated; the integrals $I(L_B)$ and $I_L$ were calculated by
numerically integrating equation~\ref{eq:ginzparam2}. The expected KTHNY scaling for liquid, hexatic, and crystal are given by\cite{nelson1979}:
\begin{equation}
G_{6,j}(r) \, \sim \,  
\begin{cases}
\exp(-r/\xi) & \text{liquid} \\ 
r^{-\eta} & \text{hexatic} \\
\text{Constant} \ne 0 & \text{crystal}
\end{cases} 
\label{eq:sca1}
\end{equation}
where $\xi$ is the correlation length and $\eta$ is the exponent characterizing the algebraic decay of the order parameter correlation function. (The exponent $\eta$ has its origins from the Ornstein-Zernikie formalism \cite{stanley1971} discussed later, however, the exponent values and functional forms for liquid-liquid critical phenomena and KTHNY formalisms are different from one-another). Equations~\ref{eq:ginzparam2} \&~\ref{eq:sca1} lead to the following scaling for $I(L)$:
\begin{equation}
I(L) \, = \,  
\begin{cases}
\text{Constant}/L^2 & \text{liquid} \\ 
L^{-\eta}/(2-\eta) & \text{hexatic} \\
\text{Constant} \ne 0 & \text{crystal}
\end{cases} 
\label{eq:sca2}
\end{equation}
Therefore, in figure~\ref{fig:corr2}, a slope of $-2$
corresponds to the liquid phase having a finite correlation length; a
slope of $-1/4$ corresponds to an algebraic decay of $G_{6,j}(r)$ with
exponent $-1/4$ and is observed for the hexatic phase near the
hexatic-crystal transition; a slope of $0$ is observed for the crystal
phase, confirming the true long-range orientational order. The scaling
properties shown in figure~\ref{fig:corr2} is further evidence for the
attainment of equilibrium in our simulations.

\paragraph*{}

The order of the phase transition was determined by examining the
dependence of the free energy barrier separating two phases as a
function of system size. In doing so, we consider only those system
sizes for which $\gamma_{\rm{GL}} \ll 1$. In order to determine the
order of the phase transition using the Lee-Kosterlitz scaling
analysis\cite{lee1990} of the free energy surface, we obtained the
Landau free energy functions at the exact transition temperatures for
the corresponding system size. This was done as follows: the Landau
free energy surface was calculated at a temperature close to the
transition, from which we calculated the grand free energies at that
temperature. Then, by numerically integrating the Clausius-Claperon
equation, ~\mbox{$d (\Omega /T)/ d(1/T) = \langle U \rangle - \mu
\langle N \rangle$}, we located the exact transition temperature. The
Landau free energy function was re-calculated at the transition
temperature by using a weighting function equal to $\exp(\beta
\Lambda[{\overline{\Psi}}_{6,j}])$ from the initial calculation. This
procedure was repeated for each system size. The Landau free energy
and the grand free energy results for the system size of $180
\sigma_{\rm{ff}}$ are provided in figure~\ref{fig:lee}. \marginpar{\fbox{Fig.~\ref{fig:lee}}} The relative
stability of the liquid, hexatic and crystalline phases are inferred
from the Landau free energy plot (figure~\ref{fig:lee}), and the exact
transition temperatures are given by the grand free energy plot
(figure~\ref{fig:lee2}). \marginpar{\fbox{Fig.~\ref{fig:lee2}}} For the bilayer adsorbed fluid, the free
energy barrier separating hexatic and crystal phases is a linear
function of system size (figure~\ref{fig:lee}a). This implies that the
mechanism of phase transition is via nucleation and is a clear
indication of a first order phase transition in the thermodynamic
limit. The critical nucleus at the transition temperature is equal to
the system size, implying a L$^{d-1}$ ($d$ being the dimensionality of
the system) dependence on system size\cite{lee1990}. In contrast to
the bilayer case, the Lee-Kosterlitz scaling of the free energy
surface for the adsorbed monolayer ($H=0.91$~nm) in
figure~\ref{fig:lee}b establishes that the transitions are continuous in the monolayer. In this case the free energy barrier
separating the liquid and the hexatic phase is independent of system
size, a clear signature of second order transition \cite{lee1990}. We
attribute this remarkable change in the mechanism of phase transition
(spinodal decomposition in the monolayer to nucleation in the bilayer)
to the interaction of defect structures across layers (see
section~\ref{sec:discuss}).

\section{Experimental Results}
\label{sec:exptres}

The DSC and DRS results for CCl$_4$ and aniline confined in ACF are
provided in figures~\ref{fig:dsc} and~\ref{fig:drs}. \marginpar{\fbox{Figs.~\ref{fig:dsc},\ref{fig:drs}}} The DSC scans
show two peaks for each fluid reminiscent of liquid-hexatic (high
temperature) and hexatic-crystal (low temperature) heat capacity peaks
associated with Kosterlitz-Thouless phase transitions. The DRS
measurements for the dielectric constant for confined aniline confirm
the existence of two phase transitions (figure~\ref{fig:drs}a). In
addition, the molecular orientational relaxation times $\tau$,
(figure~\ref{fig:drs}b) are consistent with the existence of liquid
($\tau\sim$ns), hexatic ($\tau\sim \mu$s), and crystalline
($\tau\sim$ms) phases. The transition temperatures from simulations
and DSC for CCl$_4$ are in good agreement; also, the
transition temperatures inferred from the DSC and DRS results for
aniline are in agreement (Table~II).

\paragraph*{}

The NDE measurements for confined CCl$_4$ and aniline in ACF are
provided in figure~\ref{fig:nde} \marginpar{\fbox{Fig.~\ref{fig:nde}}} and Table~II. \marginpar{\fbox{Tab~II}} The NDE signals show
signatures of divergence at the liquid-hexatic and hexatic-crystal
transition temperatures. The NDE results for the transition
temperatures for CCl$_4$ and aniline are in near quantitative agreement
with the simulation result, and also with those from DSC, and DRS (Table~II). Below, we show that the
scaling of the NDE signal with temperature is consistent with the
KTHNY theory for liquid-hexatic and hexatic-crystal transitions.

\paragraph*{}

It was empirically known that the dipolar fluctuations associated with
polarizability contributed to a positive signal for NDE. The anomalous
increase in NDE in the vicinity of the phase transition point,
reflecting the onset of long-range correlations and associated dipolar
fluctuations in the system, has been experimentally
documented\cite{chelkowski1980}. Based on a droplet model for treating
critical fluctuations by Oxtoby and Metiu\cite{oxtoby1976}, the
dipolar fluctuations near a critical point have been correlated
(linearly) with the relevant order parameter fluctuations. De
Gennes\cite{degennes1995} showed that the conformity to the droplet
model by Oxtoby and Metiu\cite{oxtoby1976} immediately leads to NDE
scaling laws that are similar to those for critical fluctuations in light
scattering experiments. The scaling law for critical opalescence in
light scattering (where the order parameter correlation function is
$g(r)$ associated with density fluctuations) is given by the familiar
expression\cite{chaikin1995},
\begin{equation}
I(\vec{q}) = \langle \sum_{\alpha, \beta =1}^{N} \, \exp (i \vec{q} \cdot \vec{x}_{\alpha})  \, \exp (i \vec{q} \cdot \vec{x}_{\beta}) \rangle \, = \, N \left[ 1 \,+ \, \int \, g(r) \, \exp (i \vec{q} \cdot \vec{r}) \, d^dr. \right]
\label{eq:nde3}
\end{equation}
In the above equation, $I(\vec{q}) = N S(\vec{q})$, is the intensity
measured in scattering experiments, $S(\vec{q})$ is the familiar
structure factor, $d$ is the number of dimensions and $N$ the total number of molecules. The double
summation is over the total number of molecules, where
$\vec{x}_{\alpha}, \vec{x}_{\beta}$ represent the coordinate vector of
the given molecule. The second equality in equation~\ref{eq:nde3}
follows from evaluating the ensemble average\cite{chaikin1995}. By
employing the extended Ornstein-Zernikie representation
\cite{stanley1971} for the density-density correlations (above $T_c$),
namely,
\begin{equation}
g(r)-1 \, \sim \, {{1} \over {r^{d -2 + \eta}}} \exp(-r / \xi), 
\label{eq:nde5}
\end{equation}
where $\xi$ is the correlation length. The integral in
equation~\ref{eq:nde3} in the limit $q \rightarrow 0$ reduces to
\begin{equation}
I(q) \sim \xi^{d-1- \eta}. 
\label{eq:nde4}
\end{equation}
The equation~\ref{eq:nde5} corresponds to Fischer's extension of the
original Ornstein-Zernikie formalism to alleviate a logarithimic
divergence of $g(r)$ for large $r$ in two-dimensions. Therefore, the
equation provides the definition for the critical exponent $\eta$. The
exponent satisfies the equality $(2 - \eta) \nu = \gamma$, for systems
with Ising symmetry, where $\nu$ (defined below) is the critical
exponent associated with the correlation length, and $\gamma$ that with
isothermal compressibility.

According to the droplet model\cite{oxtoby1976}, the analogous
expression for NDE is given by,
\begin{equation}
\rm{NDE} \sim \int \, \langle \Phi(0) \Phi(\vec{r}) \rangle \exp (i \vec{q} \cdot \vec{r}) \, d^dr,
\label{eq:nde2}
\end{equation}
where, ~\mbox{$\langle \Phi(0) \Phi(\vec{r}) \rangle$} is the order
parameter correlation function and $d$ the dimensionality of the
system. The rationale behind the droplet model is that the NDE signal
results from the scattered intensity caused by the effective
dipole-dipole correlations in the system, the same way light
scattering intensity results from density-density correlations.

As shown by de Gennes and Prost\cite{degennes1995}, the droplet model leads to
the experimentally observed scaling for NDE in the vicinity of a
liquid-liquid critical point in three dimensions, namely,
~\mbox{$\rm{NDE} \sim \xi^{d-1- \eta}$}, with ~\mbox{$\xi= \xi_o
t^{-\nu}$}, and $t=|T-T_c|/T_c$. The theoretically predicted NDE
scaling for this system (Ising symmetry, d=3) is therefore,
\begin{equation}
\begin{cases}
t^{- (2 - \eta) \nu} & d=3, \text{Ising}; \hspace{1cm} \text{e.g., liquid-liquid} \\
t^{-1.25} & \nu=0.64, \eta=0.041 ; \hspace{1cm}  \text{3-d Ising model} \\
t^{-1.28} & \nu=0.64, \eta=0 ; \hspace{1cm}  \text{Ornstein-Zernikie} \\
t^{-1} & \nu=0.5, \eta=0 ; \hspace{1cm}  \text{Mean field theory} \\
t^{-1} & \text{dipolar fluids} ; \hspace{1cm}  \text{experiment \cite{chelkowski1980}.} \\
\end{cases} 
\label{eq:nde10}
\end{equation}

\paragraph*{}

Encouraged by the agreement of the phenomenological NDE scaling laws
with experiment for systems with Ising symmetry, we derive the
analogous NDE scaling laws near the liquid-hexatic, and
hexatic-crystal phase boundaries. Our analysis was greatly facilitated
because Halperin and Nelson\cite{nelson1979} and
Young\cite{nelson1979} have worked out the scaling of the relevant order
parameter correlations to be used in equations~\ref{eq:nde3}
and~\ref{eq:nde4}. The order parameter correlations above the
transition temperatures for the liquid-hexatic and hexatic-crystal
transitions have the form, ~\mbox{$G_{T,6}(r) \sim \exp (-r / \xi_{+})$}, where
the subscripts $T,6$ correspond to translational and orientational
correlation functions, and $+$ refers to the correlation function above
the transition temperature, i.e., $t>0$. Above the hexatic-liquid phase
transition temperature, ~\mbox{$\xi_{+} = A \exp(B / |t|^\nu)$}, with
$\nu=0.5$ based on a renormalization group analysis. Similarly, the
translational order parameter correlation length is given by ~\mbox{$\xi_{+}
= A \exp(B / |t|^\nu)$}, with $\nu=0.37$ for a smooth substrate, above
the crystal-hexatic phase boundary. The phenomenological NDE scaling
directly follows from using equation~\ref{eq:nde4} (note that
$\eta=0, d=2$),
\begin{equation}
1/\rm{NDE} = A \exp(-B/|T-T_c|^{\nu})
\label{eq:nde6}
\end{equation}
with $\nu=0.5$ for the
liquid-hexatic (high temperature) transition and $\nu=0.37$ for
the hexatic-crystal (low temperature) transition.

\paragraph*{}

The NDE scaling with temperature for confined aniline is re-plotted in
figure~\ref{fig:ndescaling} \marginpar{\fbox{Fig.~\ref{fig:ndescaling}}} along with the NDE scaling law
(equation~\ref{eq:nde6}). The results show the consistency of the
NDE signals with KTHNY scaling laws for the liquid-hexatic and
hexatic-crystal transitions. The estimates for the exponent $\nu$ for
liquid-hexatic and hexatic-crystal transitions based on the
experimental measurements is currently not possible owing to the
nature of the ACF sample; true divergence is not observed in our
measurements as the length of the in-plane correlations in the fluid
are curtailed by the finite size of the graphitic microcrystals in
ACF. Therefore, NDE signals close to the transition temperatures for
an infinite system are not accessible in our experiments. Instead, the
NDE signals in our experiment correspond to a finite system size of
5~nm, the average size of the graphitic microcrystal (in the $xy$
plane) in our ACF-A10 sample.

\section{Discussion}
\label{sec:discuss}

Our computer simulations for a bilayer of adsorbed fluid in a
slit-pore with smooth walls has established the existence of two-stage
melting in the thermodynamic limit via system size scaling
analysis. Based on the calculations of the Ginzburg parameter we
establish that for system sizes smaller than 60 molecular diameters,
fluctuations are too large to uphold the finite size simulation
results. Based on the Lee-Kosterlitz scaling of the free energy
surface, we establish that the transitions are first order in the
thermodynamic limit. Similar results for a monolayer established that the
transitions were continuous. Our experimental results rely on indirect
evidence based on phase transitions (structural measurements are
considered direct evidence).  Nevertheless, the quantitative agreement
between the simulations and DSC, DRS, and NDE measurements, make a strong
case for the existence of KTHNY melting and therefore a hexatic phase
in the confined system. This interpretation, if correct, shows that
the hexatic phase is stable over a wide temperature range, 55~K for
CCl$_4$ and 26~K for aniline. This large range of state conditions
over which the hexatic phase is stable may be unique to confined
fluids in porous media.

\vspace{\baselineskip}

\noindent \textbf{Effect of the Six-Fold Potential.} 
The main difference between the simulation results and the
experimental results reported in this paper is that the former
corresponds to a smooth substrate, while the latter corresponds to a
substrate with 6-fold symmetric potential. On a theoretical basis, the
effects of the six-fold symmetry of the pore potential and that of the
finite size of the graphitic microcrystals combined might be expected to
nullify the liquid-hexatic phase transition and cause a rounding of
divergences associated with the hexatic-crystal transitions. The
latter is observed in the NDE measurements (the NDE signals do not
diverge), while the former manifests itself as a remnant KTHNY behavior,
presumably due to a weak six-fold substrate potential. It is worth
pointing out that the LJ diameters of CCl$_4$ and aniline (0.5-0.6~nm)
are much larger than the C-C bond length in graphite (0.14~nm), so
that the fluid molecules only feel a mild corrugation in the
fluid-wall potential in passing along the surface. This feature
contrasts with the situation for earlier experiments concerning Xe,
Ar, and Kr on graphite substrates, where the LJ diameters for the
fluid are much smaller and comparable to the C-C spacing in graphite.
There is evidence of an intrinsic hexatic phase even for the case of
Xe in graphite\cite{rosenbaum1983,greiser1987}; therefore, for substrate
molecules larger than Xe, we expect the crystal to melt into a hexatic
phase with intrinsic stiffness with respect to bond-orientational
fluctuations. While we note the quantitative agreement in transition
temperatures between simulations and experiments and have described
the respective domains of validity, the question of whether an
intrinsic hexatic phase is observed in a realistic simulation with
six-fold symmetric substrate potential remains to be
answered. Simulations are currently in progress to address this
question. On the experimental side, the challenge is to synthesize a
porous sample with uniform pores with large microcrystal sizes to
examine the scaling for several decades in $|T-T_c|/T_c$ on a log
scale, and to obtain direct structural evidence of the hexatic phase
via scattering experiments.

\vspace{\baselineskip}

\noindent \textbf{Effect of Multi-Layers on KTHNY Melting.} The
scaling of the order-parameter correlation functions in the
simulations (figure~\ref{fig:corr2}) are consistent with the KTHNY
behavior, implying that it is the vortex excitations that govern the
equilibrium behavior in the quasi-two-dimensional systems and that the
melting transition is defect-mediated. Moreover, for a
quasi-two-dimensional monolayer, the Kosterlitz-Thouless transitions
are continuous\cite{radhakrishnan2002a}, while for quasi-two-dimensional
bilayers the Kosterlitz-Thouless transitions become first-order
(figure~\ref{fig:lee}). We ascribe this deviation from 2-d behavior to
the interactions between the defect configurations in different layers,
based on the following scaling arguments.

\paragraph*{}

On heuristic grounds\cite{kosterlitz1972}, the free energy of exciting a vortex pair of opposite winding numbers
relative to the ordered phase in the $xy$ model in 2-d is given by
~\mbox{$\Delta F = (\pi J - 2 k_B T) \log
(L)$}, $J$ being the interaction energy of two
neighboring spins with the same alignment and $L$ being the distance
between the vortex cores. This is the essence of the
Kosterlitz-Thouless~(KT) transition.  If we consider two planes of
$xy$ models interacting with each other, and the vortices in one layer
are perfectly in alignment with those in the second layer, the free
energy relative to the ordered phase is given by ~\mbox{$\Delta
F^{(\rm{KT})} = (2 \pi J - 2 k_B T) \log (L)$}, which is
qualitatively the same as the KT behavior. This configuration,
however, corresponds to a reduced entropy situation because the number
of different ways of placing the aligned vortex pairs is the same as
the single layer case. A second higher entropy scenario exists,
where a vortex in one layer is aligned with a vortex of the opposite
winding number in the other layer, with the cores displaced by
distance $A$. Such a situation corresponds to a free energy (see Appendix): \marginpar{\fbox{Appendix}}
\begin{equation}
\Delta F =  \Delta F^{\rm{KT}} \,+ \, (\pi-1)J' L^2 \,+\,  J'AL \ln L -k_BT \ln (1 + \pi A^2)
\label{eq:intrvortex}
\end{equation}
where $J$ is the interaction of nearest neighbors of spins with the
same alignment, $L$ is the in-plane distance between vortex cores,
~\mbox{$\Delta F^{(\rm{KT})} = (2 \pi J - 2 k_B T) \log (L)$} is the
Kosterlitz-Thouless free energy\cite{kosterlitz1972}, $J'$ is the
nearest-neighbor interaction between layers and in general can be
different from $J$, and $A$ is the ``offset-distance'' between vortices as defined above. Treating $L$ as the order parameter, the free
energy profile is obtained by the locus of points that minimize
$\Delta F$, the variational parameter being $A$. The locus of points
minimizing $\Delta F$ is qualitatively different from that for the true KT
behavior, given by $\Delta F^{(\rm{KT})}$ (figure~\ref{fig:vort}) \marginpar{\fbox{Fig.~\ref{fig:vort}}}. In
particular, at small separations of the vortices, the bound state
actually exists as a metastable state for $T>T_c$, a clear signature
of a first-order phase transition.  Figure~\ref{fig:vort} depicts the
free energy profile for the monolayer case and bilayer case at a
temperature $k_B T=1.5$, assuming $2 \pi J=1$ and $(\pi-1)J'=0.2$
(where $J,J'$ are in units of $k_BT$). The plot marked by the solid
line represents $\Delta F^{(\rm{KT})}$, the true KT scenario, at a
temperature, $T$, greater than the transition temperature, $T_c$, for
which the lowest free energy state corresponds to $L=\infty$, i.e.,
the vortices are unbound. Moreover, the bound state ($L=1$) is
unstable and remains so, as long as $T>T_c$, and therefore the transition is
continuous (second-order). The set of dotted lines are plotted
according to equation~(\ref{eq:intrvortex}) for different values of
the offset distance $A$. It is immediately clear that the locus of
points minimizing $\Delta F$ is qualitatively different from the KT
plot (equation for $\Delta F^{(\rm{KT})}$). In particular, at small
separations of the vortices, the bound state actually exists as a
metastable state for $T>T_c$, a clear sign of a first-order phase
transition. In other words, the vortex excitations that lead to
equation~\ref{eq:a4} (see Appendix) imply a pathway for nucleation of the ordered
phase, i.e., by the annihilation of vortices $\vec{U}^{+}$ and
$\vec{U}^{-}$ across layers. This is possible only if hopping of
particles between layers are allowed, such as in our confined fluid
system.

\paragraph*{}

The simulations and above arguments pertain to smooth pore walls.
Therefore, the effect of the six-fold substrate potential was ignored
while taking into account the strong potential due to the porous
matrix. To our knowledge, this is the first report of the effect of
multi-layers on KTHNY behavior.

\vspace{\baselineskip}

\noindent \textbf{Effect of Strength of the Pore Potential.} It was remarked
earlier, based on the evidence from simulations and experiments, that
the hexatic phase was stable over a large temperature range for
CCl$_4$ and aniline. The ratio of the strength of the fluid-wall
interaction to the fluid-fluid interaction, $\alpha$
(equation~\ref{eq:alpha}), has a significant effect on the melting
phase diagram\cite{radhakrishnan2002} (see figure~\ref{fig:glob}); \marginpar{\fbox{Fig.~\ref{fig:glob}}} in
particular, the larger the value of $\alpha$, the larger the temperature
range over which the hexatic phase is stable. Therefore, fluids with
purely dispersive interactions (e.g., CCl$_4$, methane, benzene etc.)
confined in ACF, with large values of $\alpha$ show a large
temperature range of stability for the hexatic phase. For strongly
dipolar fluids, $\alpha$ is smaller, the temperature range of
stability of the hexatic phase decreases\cite{radhakrishnan2002}, and may vanish for sufficiently small $\alpha$.

\paragraph*{}

We acknowledge funding from NSF under grant number CTS-0211792 and
from KBN under grant number 2~PO3B~01424. The experiments were
performed in Adam Mickiewicz University, Poznan, Poland. The
supercomputing time was provided by the San Diego Supercomputing
Center under a NRAC grant (number MCA93S011P). International
cooperation was supported by a NATO Collaborative Linkage grant
(no. 978802).

\newpage
\renewcommand{\theequation}{A-\arabic{equation}}
\setcounter{equation}{0}  

\vspace{12pt}

\section*{Appendix}

We give here the derivation of equation~\ref{eq:intrvortex} for a bilayer of the $xy$ model of spins on a lattice. We consider two planes of
$xy$ models interacting with each other; vortices with opposite winding numbers  in one layer
are perfectly in alignment with those in the second layer (i.e, the vortex cores span two layers). The free energy relative to the ordered phase is given by\cite{kosterlitz1972}:
\begin{equation}
\Delta F^{(\rm{KT})} = (2 \pi J - 2 k_B T) \log (L)
\label{eq:a1}
\end{equation}
which is qualitatively the same as the KT behavior (solid line in figure~\ref{fig:vort}). This configuration,
however, corresponds to a reduced entropy situation because the number
of different ways of placing the aligned vortex pairs is the same as
the single layer case. A second, higher entropy scenario exists,
where a vortex in one layer is aligned with a vortex of the opposite
winding number in the other layer, with the cores displaced by
distance $A$.

\paragraph*{}

Vortices of positive and negative winding numbers whose cores are displaced by length $A$ are represented as vector fields $\vec{U}^{+}$ and $\vec{U}^{-}$, given by:
\begin{eqnarray}
\vec{U}^{+} \,=\, (x -A)\hat{e}_x \,-\, y \hat{e}_y \\
\vec{U}^{-} \,=\, -y\hat{e}_x \,+\, (x -A)\hat{e}_y
\end{eqnarray}
where ${e}_x,{e}_y$ are unit vectors in the $x$ and $y$ directions. The interaction energy $E$ relative to the ordered phase, for the given spin fields $\vec{U}^{+}$ and $\vec{U}^{-}$ in each layer, is given by: 
\begin{eqnarray}
E\,=\, J' \, \int_0^L \int_0^L \, \vec{U}^{+} \cdot \vec{U}^{-} \, dx \,  dy \, + \pi J'L^2 \\
 \,=\, (\pi-1)J' L^2 \,+\, J'AL \ln L
\label{eq:a3}
\end{eqnarray}
and the excess entropy associated with placing the cores with an offset distance $A$ is given by $S = k_B \ln (1+ \pi A^2)$. The free energy relative to the ordered state is given by:
\begin{eqnarray}
F \,=\, 2\pi J \ln L \,-\, 2k_BT \ln L \,+\, (\pi-1)J' L^2 \,+\, J'AL \ln L -k_BT \ln (1 + \pi A^2) \\
  \,=\, \Delta F^{\rm{KT}} \,+ \, (\pi-1)J' L^2 \,+\,  J'AL \ln L -k_BT \ln (1 + \pi A^2)
\label{eq:a4}
\end{eqnarray}

\newpage
\begin{figure}
\caption{Orientational correlation functions for the confined
bilayer of adsorbate ($H=1.41$~nm): liquid (360~K), hexatic (335~K),
and crystal (290~K) phases.}
\label{fig:corr}
\end{figure}

\begin{figure}
\caption{Scaling behavior of orientational correlation functions. The solid line corresponds to zero slope, the dashed line corresponds to a slope of $-1/4$ and the dotted line corresponds to a slope of $-2$.}
\label{fig:corr2}
\end{figure}

\begin{figure}
\caption{(a) Landau free energy functions for two different system
sizes $60 \sigma_{\rm{ff}}$ and $180 \sigma_{\rm{ff}}$ for a pore
width of $H=1.41$~nm at the hexatic-crystal transition. The
temperature was $T=293$~K for the $60 \sigma_{\rm{ff}}$ system, and
$T=290$~K for the $180 \sigma_{\rm{ff}}$ system. (b) Corresponding results for $H=0.91$~nm system at the liquid-hexatic transition, for the $60 \sigma_{ff}$ system ($T=390$~K) and $180 \sigma_{ff}$ ($T=387$~K).}
\label{fig:lee}
\end{figure}

\begin{figure}
\caption{The grand free
energy per mole for liquid, hexatic and crystal phases ($H=1.41$~nm); the
cross-over points correspond to phase transitions and give the
transition temperatures (see Table~II).}
\label{fig:lee2}
\end{figure}

\begin{figure}
\caption{DSC scan for CCl$_4$ and aniline confined in activated carbon
  fiber ACF~A-10 at a temperature scanning rate of $0.1$~K$/$min. The peaks are interpreted as the liquid-hexatic (high $T$) and hexatic-crystal (low $T$) transitions. The approximate transition temperature is located at the tail of each peak in the low temperature side (see Figure~9.4.3 on page~550 of Chaikin and Lubensky\cite{chaikin1995}).}
\label{fig:dsc}
\end{figure}

\begin{figure}
\caption{DRS temperature scan (the sample was equilibrated at each
temperature) for aniline confined in ACF-A10, as a function of
temperature indicating phase transitions (dashed lines), (a)
capacitance; (b) molecular orientational relaxation time ($\tau$). The
disappearance of the relaxation branch due to conductance of the
sample is believed to correspond to the liquid-hexatic transition, and
the change in the relaxation time of the confined fluid is interpreted
as the occurrence of the hexatic-crystal phase
transition\cite{sliwinska2001}. The relaxation branch corresponding to
the Maxwell-Wagner effect occurs as a result of signal dispersion due to
the suspended ACF particles in bulk aniline\cite{sliwinska2001}.}
\label{fig:drs}
\end{figure}

\begin{figure}
\caption{NDE for (a) CCl$_4$, and (b) aniline, confined in activated
  carbon fiber ACF~A-10. The peaks correspond to the liquid-hexatic
  (L/H) and hexatic-crystal (H/C) transitions. Lines are drawn through
  the points as guide to the eye.}
\label{fig:nde}
\end{figure}

\begin{figure}
\caption{NDE scaling for aniline confined in activated carbon fiber
ACF~A-10. The filled circles correspond to liquid above the L/H phase
transition, and the filled squares correspond to the hexatic above the
H/C phase transition. The solid lines represent a fit to the scaling
law with the theoretically predicted exponents (see legend). For the
liquid (filled circles) $A=\exp(34.52)$, $B=6.48$, $\nu=0.5$, $T_c =
42^{\circ}$C; for the hexatic (filled squares) $A=\exp(32.46)$,
$B=0.338$ , $\nu=0.37$, $T_c = 27^{\circ}$C.}
\label{fig:ndescaling}
\end{figure}

\begin{figure}
\caption{Free energy scaling for a set of interacting vortex pairs in
a bilayer according to the equation for $\Delta F^{(\mbox{KT})}$ and
equation~\ref{eq:intrvortex}. The distances $L$ and $A$ are in
dimensionless units (scaled by the lattice length) and the free energy
is in units of $k_BT$.}
\label{fig:vort}
\end{figure}

\begin{figure}
\caption{Global phase diagram of a fluid in a slit pore of width $H=3
  \sigma_{\rm{ff}}$ from simulations and free energy
  calculations\cite{radhakrishnan2002,radhakrishnan2002a} (open
  symbols) and experiment (filled
  symbols)\cite{radhakrishnan2002,radhakrishnan2002a,sliwinska-bartkowiak2001,radhakrishnan2000,sliwinska2001,kaneko_psd,watanabe1999}. Three
  different phases are observed: liquid~(L), hexatic~(H), and
  crystalline~(C). The dashed line represents an extrapolation of the
  phase boundaries based on MC simulations without free energy
  calculations. The simulations are for a LJ fluid in slit-pore with
  different values of $\alpha$. The experiments are for various
  adsorbates confined within activated carbon fibers (ACF, mean pore
  width 1.4~nm): H$_2$O\cite{sliwinska-bartkowiak2001} ($\alpha=0.51$),
  C$_6$H$_5$NO$_2$\cite{radhakrishnan2000} ($\alpha=1.22$),
  C$_6$H$_5$NH$_2$\cite{sliwinska2001} ($\alpha=1.75$),
  CH$_3$OH\cite{sliwinska-bartkowiak2001} ($\alpha=1.82$),
  CCl$_4$\cite{kaneko_psd,radhakrishnan2002a} ($\alpha=1.92$),
  C$_6$H$_6$\cite{watanabe1999} ($\alpha=2.18$).}
\label{fig:glob}
\end{figure}

\newpage
\begin{center}
\epsfig{file=bigor.eps,width=4.0in,angle=0}
\end{center}

\newpage
\begin{center}
\epsfig{file=ginz.eps,width=3.0in,angle=-90}
\end{center}

\newpage
\begin{center}
\epsfig{file=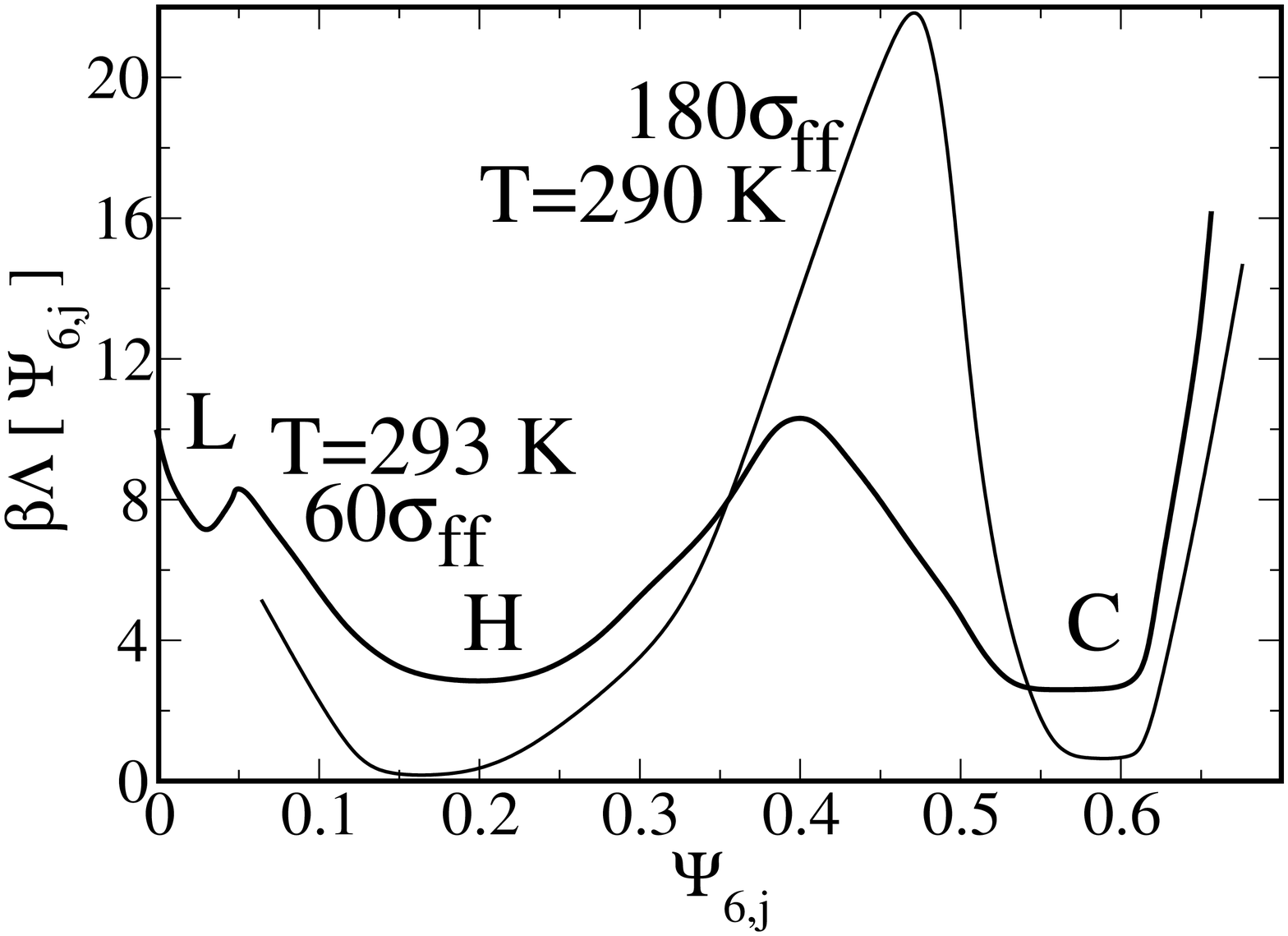,width=4.0in,angle=-0}

\vspace{12pt}

(a)

\vspace{24pt}

\epsfig{file=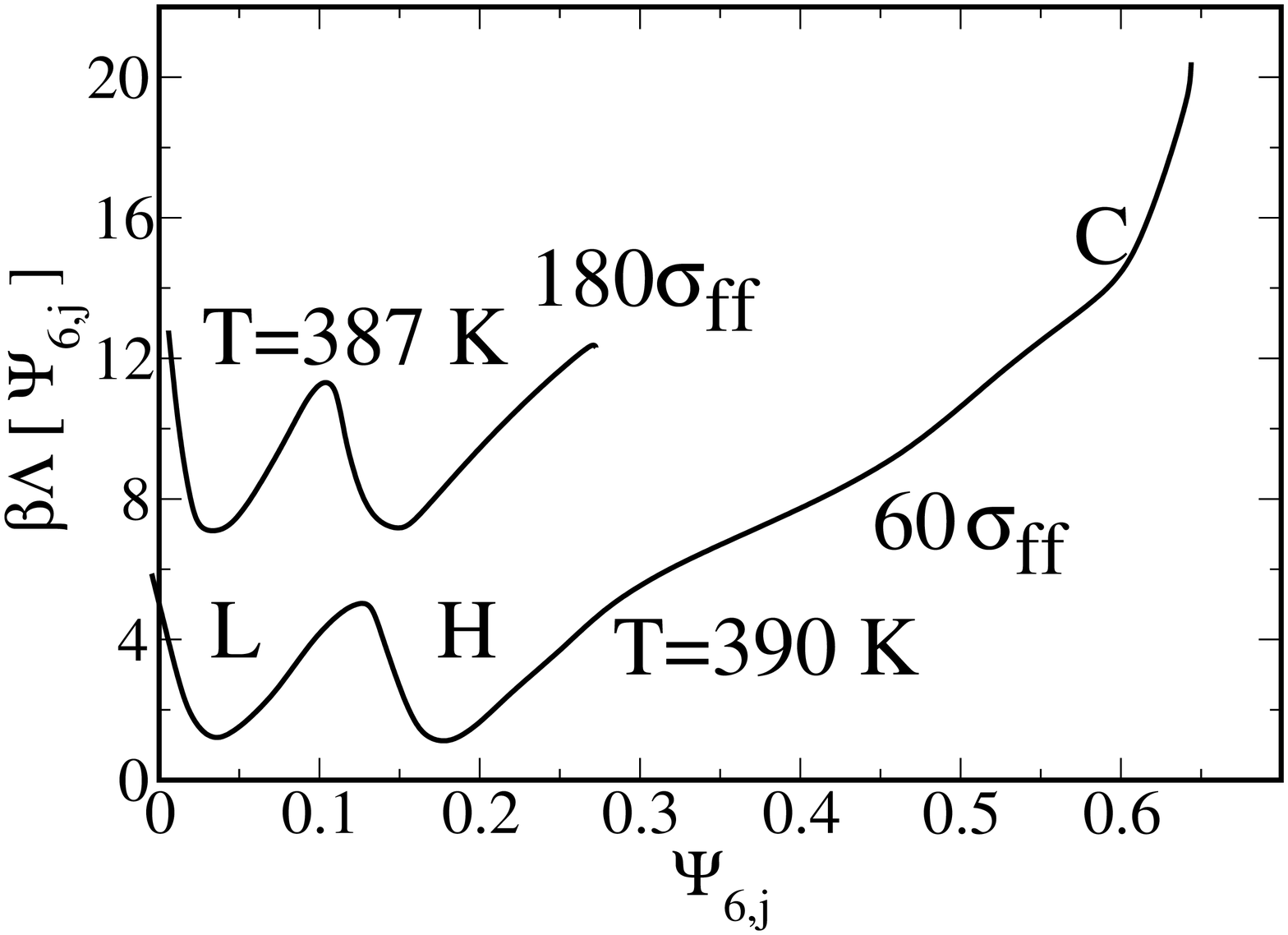,width=4in,angle=0}

\vspace{12pt}

(b)
\end{center}

\newpage
\begin{center}
\epsfig{file=hex_gfe.eps,width=4.0in,angle=0}
\end{center}

\newpage
\begin{center}
\epsfig{file=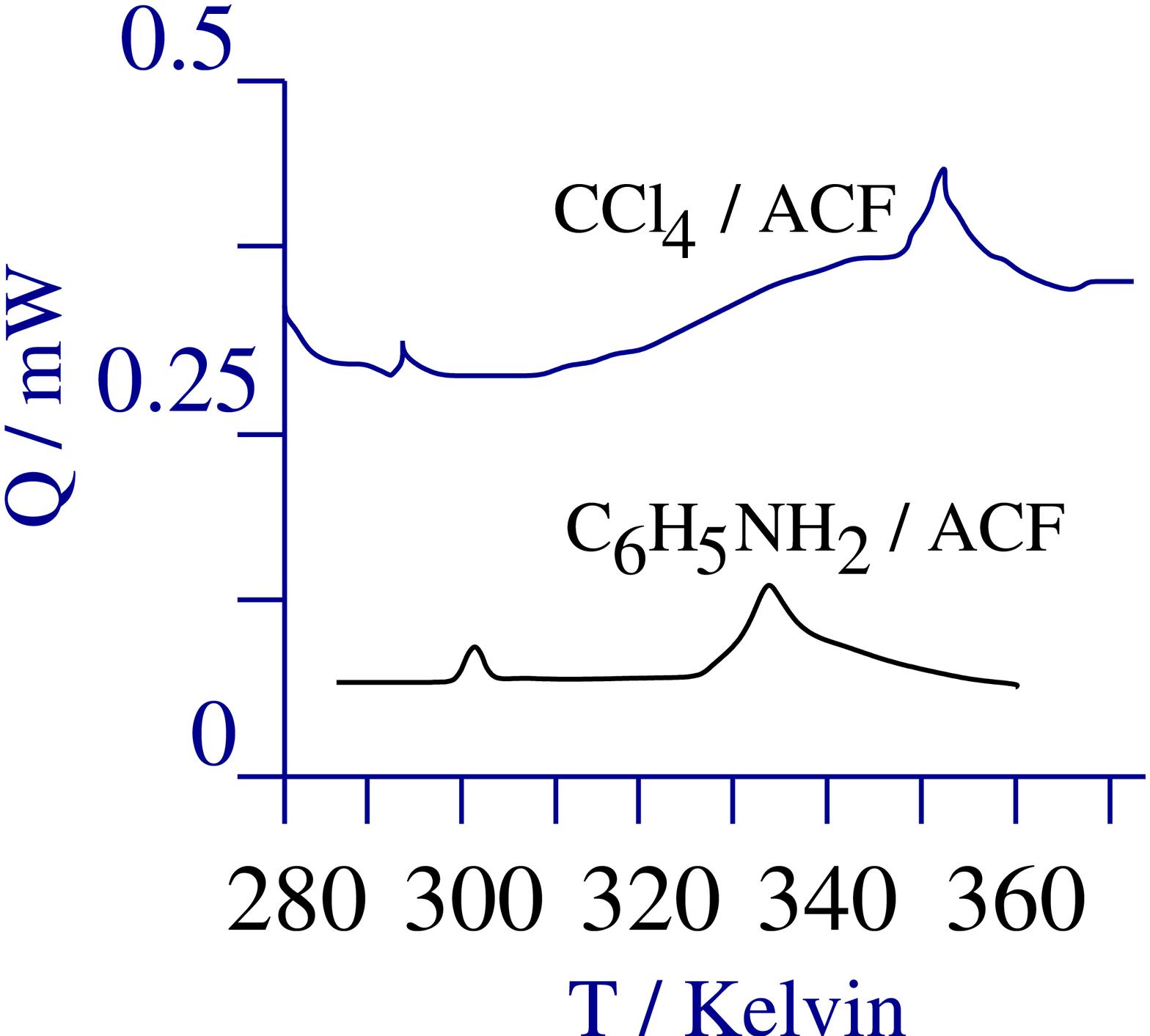,width=4.0in,angle=0}
\end{center}

\newpage
\begin{center}
\epsfig{file=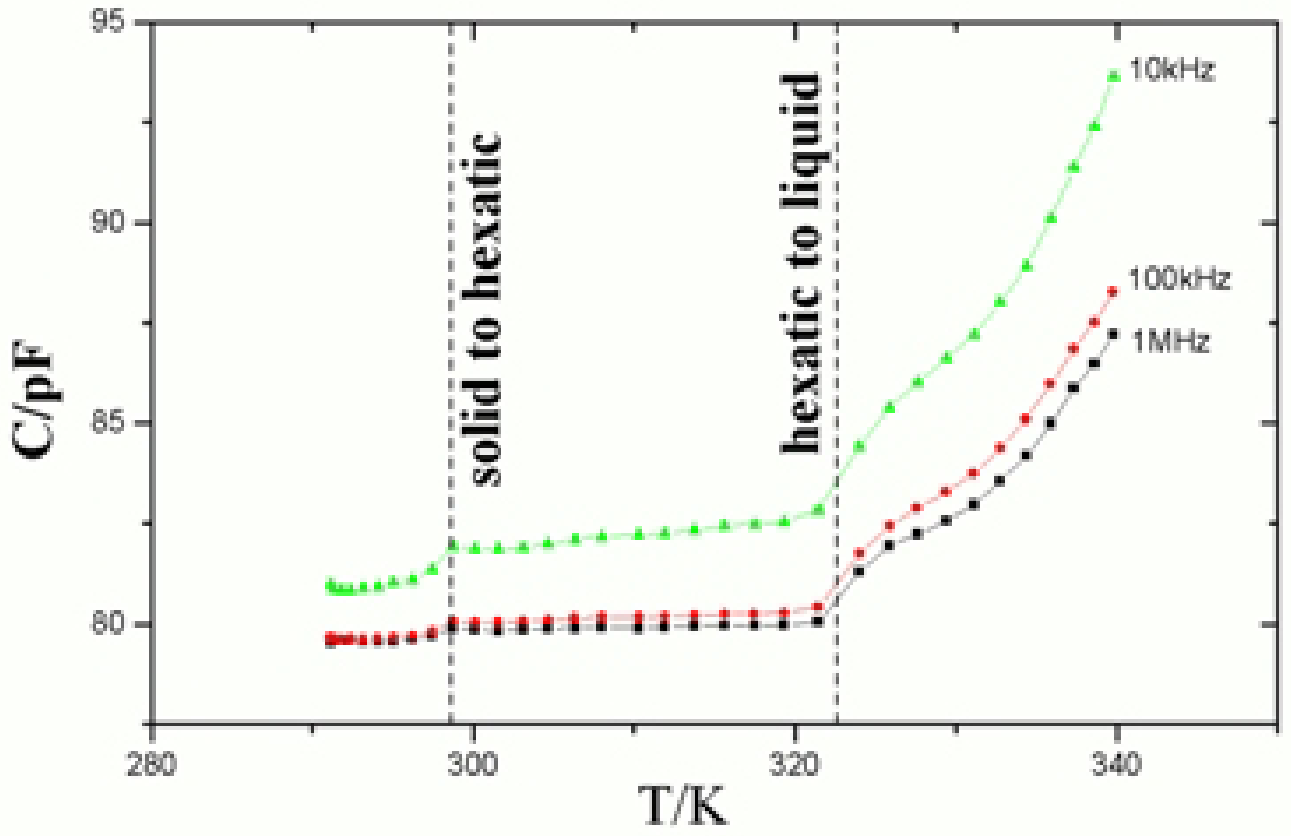,width=4.0in,angle=0}

\vspace{12pt}

(a)

\vspace{24pt}

\epsfig{file=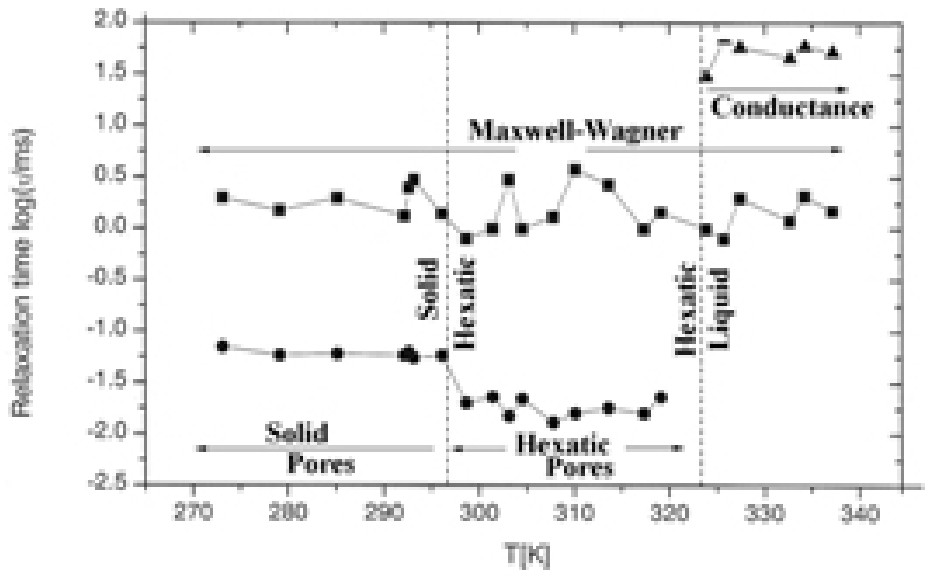,width=4.0in,angle=0}

\vspace{12pt}

(b)

\end{center}

\newpage
\begin{center}
\epsfig{file=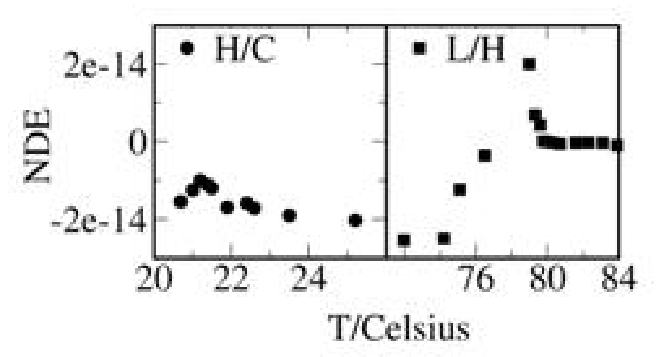,width=4.0in,angle=0}

\vspace{12pt}

(a)

\vspace{24pt}

\epsfig{file=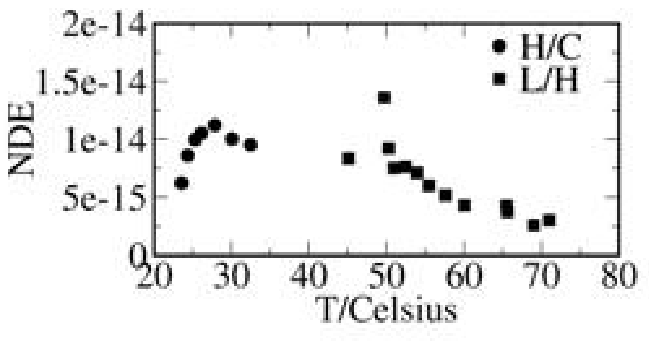,width=4.0in,angle=0}

\vspace{12pt}

(b)

\epsfig{file=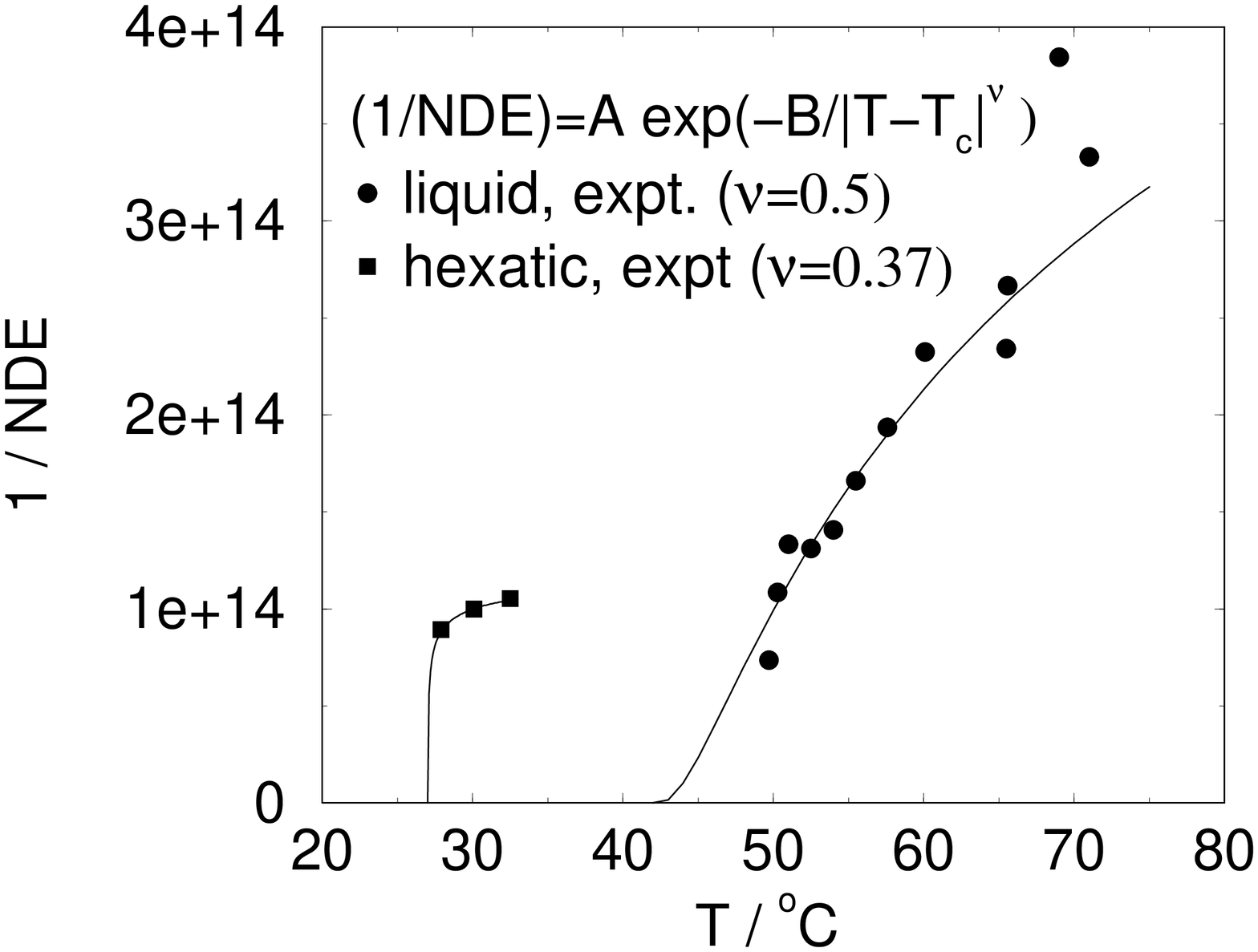,width=4.0in,angle=-90}
\end{center}

\newpage
\begin{center}
\epsfig{file=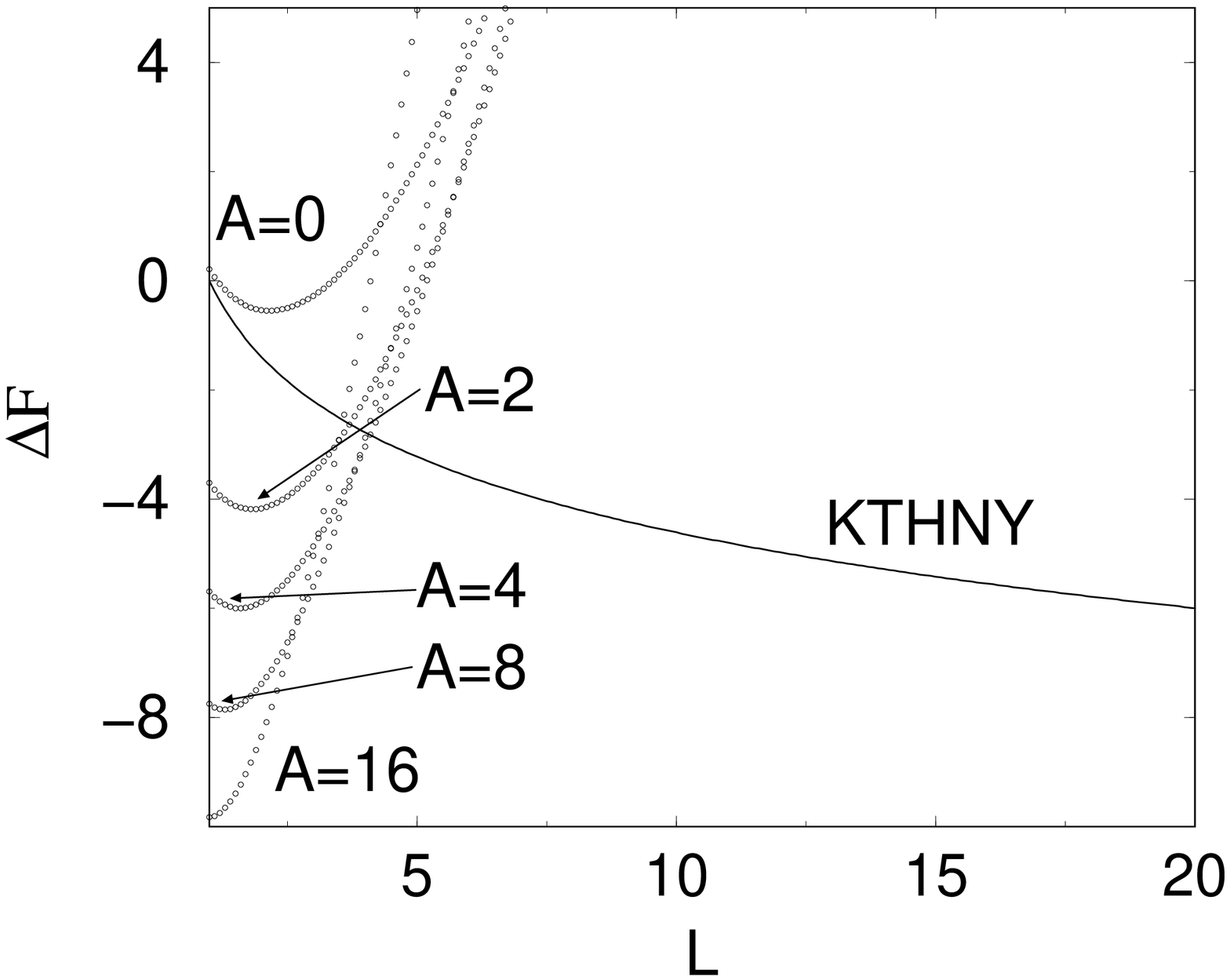,width=4.0in,angle=-90}
\end{center}    

\newpage
\begin{center}
\epsfig{file=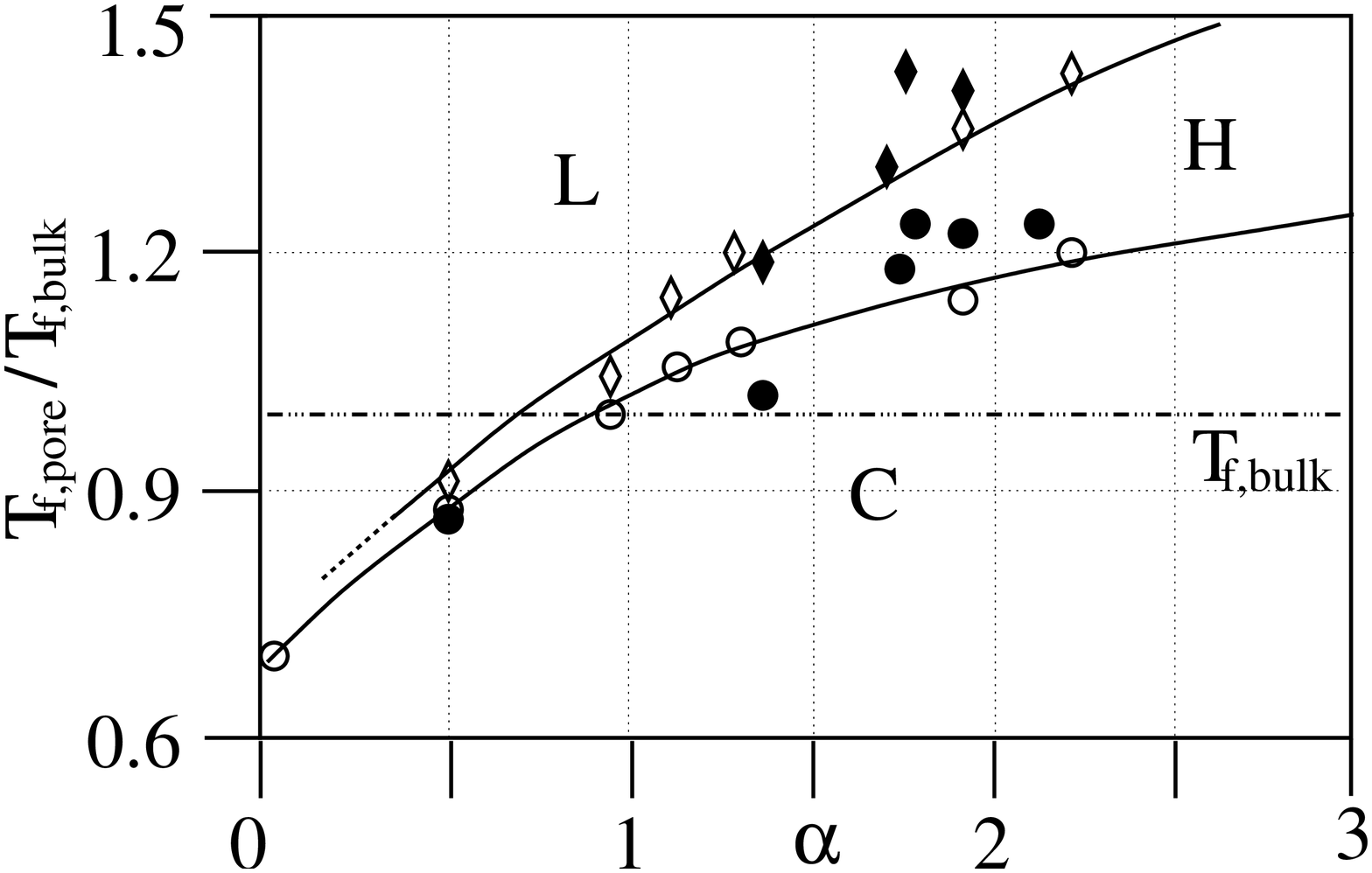,width=4.0in,angle=0}
\end{center} 

\newpage
\begin{tabular}{c c c c c}

\multicolumn{5}{c}{Table I: Ginzburg parameter}\\
\multicolumn{5}{c}{for different $L$ and $T$.}\\
\hline
\hline
$T/$K & $10 \sigma_{\rm{ff}}$ &  $40 \sigma_{\rm{ff}}$ &   $60 \sigma_{\rm{ff}}$ &  $180 \sigma_{\rm{ff}}$ \\
\hline
340 & 8.98 & 1.95 & 0.117 & 0.038  \\
330 & 6.4  & 1.7  & 0.102 & 0.029  \\
320 & 4.29 & 1.17 & 0.098 & 0.019  \\
310 & 3.11 & 0.97 & 0.16  & 0.0189 \\
300 & 3    & 0.937& 0.176 & 0.0189 \\
\hline
\end{tabular}

\vspace{48pt}

\begin{tabular}{|c|c|c|c|c|}
\multicolumn{5}{c}{}\\
\multicolumn{5}{c}{Table II. Transition temperatures from simulation and}\\
\multicolumn{5}{c}{experiment for CCl$_4$ and aniline in ACF-10.} \\
\hline
\hline
Fluid & \multicolumn{2}{|c|}{Low $T_c/$K (H/C)} & \multicolumn{2}{|c|}{High $T_c/$K (L/H)} \\ \cline{2-5}
 & Simulation & Expt. & Simulation & Expt. \\
\hline
CCl$_4$ & 290 & 298$^{a}$, 295$^{b,c}$ & 348 K & 348$^a$, 353$^b$, 352$^c$ \\
Aniline & -- & 301$^a$, 298$^{b}$, 300$^c$ & --  & 325$^a$, 324$^{b}$, 315$^c$ \\
\hline
\multicolumn{5}{l}{$^a$DSC,  $^b$DRS,  $^c$NDE} \\
\multicolumn{5}{c}{}\\
\end{tabular}

\end{document}